# State-dependent modulation of locomotion by GABAergic spinal sensory neurons

Kevin Fidelin[1,2,3,4], Lydia Djenoune[1,2,3,4,5], Caleb Stokes[1,2,3,4], Andrew Prendergast[1,2,3,4], Johanna Gomez[1,2,3,4], Audrey Baradel[1,2,3,4], Filippo Del Bene[4,6], and Claire Wyart[1,2,3,4,*]

[1]Institut du Cerveau et de la Moelle épinière (ICM), F-75013, Paris, France
[2]INSERM UMRS 1127
[3]CNRS UMR 7225
[4]UPMC Univ Paris 06, F-75005, Paris, France
[5]Museum National d'Histoire Naturelle, F-75005, Paris, France
[6]Institut Curie, CNRS UMR 3215, INSERM U934, F-75005, Paris, France

[*] Corresponding author: claire.wyart@icm-institute.org

**Running title**

Modulation of locomotion by CSF-contacting neurons

**Summary**

The cerebrospinal fluid (CSF) constitutes an interface through which chemical cues can reach and modulate the activity of neurons located at the epithelial boundary within the entire nervous system. Here, we investigate the role and functional connectivity of a class of GABAergic sensory neurons contacting the CSF in the vertebrate spinal cord, and referred to as CSF-cNs. The remote activation of CSF-cNs was shown to trigger delayed slow locomotion in the zebrafish larva, suggesting that these cells modulate components of locomotor central pattern generators (CPGs). Combining anatomy, electrophysiology, and optogenetics *in vivo*, we show that CSF-cNs form active GABAergic synapses onto V0-v glutamatergic interneurons, an essential component of locomotor CPGs. We confirmed that activating CSF-cNs at rest induced delayed slow locomotion in the fictive preparation. In contrast, the activation of CSF-cNs promptly inhibited ongoing slow locomotion. Moreover, selective activation of rostral CSF-cNs during ongoing activity disrupted rostrocaudal propagation of descending excitation along the spinal cord, indicating that CSF-cNs primarily act at the premotor level. Altogether, our results demonstrate how a spinal GABAergic sensory neuron can tune the excitability of locomotor CPGs in a state-dependent manner by projecting onto essential components of the excitatory premotor pool.

**Introduction**

During active locomotion, sensory afferent neurons provide excitatory feedback to motor neurons and spinal interneurons in response to muscle contraction. Local GABAergic interneurons can modulate this pathway by inhibiting sensory afferents at the presynaptic level [1, 2]. Genetic targeting and manipulation of these GABAergic interneurons recently demonstrated the importance of presynaptic modulation of sensory afferents to control fine motor behaviors in mice [3, 4]. Although GABAergic modulation is essential for controlling excitability and spike timing of excitatory neurons throughout the nervous system [5, 6], little is known about the GABAergic modulation of descending premotor excitatory interneurons controlling rhythm and pattern generation in the spinal cord. Pharmacological manipulations showed that GABAergic neurons could modulate the burst frequency of motor neurons during fictive locomotion [7, 8], suggesting that the release of GABA can control the excitability of spinal excitatory interneurons driving fictive locomotion. Yet, the nature of GABAergic neurons mediating this effect and their targets in the spinal cord remain to be identified.

Almost a century ago, Kolmer and Agduhr identified cerebrospinal fluid-contacting neurons (CSF-cNs), also called KA cells in African clawed frog and zebrafish, as villiated neurons surrounding the central canal in the spinal cord of over 200 vertebrate species [9–11]. Spinal CSF-cNs are sensory neurons with unique features because they are GABAergic, intraspinal, and reside in the ventral part of the spinal cord. As CSF-cNs exhibit longitudinal axons projecting in the ventral cord, they might relay chemical information from the CSF to spinal circuits [12–14]. Despite the conservation of CSF-cNs among vertebrates, their role in sensorimotor integration is still poorly understood. Remote activation of CSF-cNs was shown to trigger delayed slow locomotor activity in head-restrained zebrafish larvae, indicating that CSF-cNs could project onto components of the slow swimming central pattern generator (CPG) [14]. However, the nature of postsynaptic targets of CSF-cNs within the slow locomotor CPG remains to be identified.

Here, we took advantage of the zebrafish larva to optically probe the cellular and circuit mechanisms deployed by GABAergic CSF-cNs neurons to modulate locomotor activity in an intact animal. We generated a specific line to analyze the morphology of CSF-cNs, manipulate their activity, and map their functional connectivity onto locomotor CPGs. We performed Channelrhodopsin-2 (ChR2)-mediated activation of CSF-cNs in combination with whole cell recordings of their targets to demonstrate that CSF-cNs form GABAergic synapses onto $dbx1^+/evx1^+$ glutamatergic commissural descending V0-v interneurons. These interneurons are essential components of the locomotor CPG in multiple vertebrate species [15–19], and are selectively active during slow locomotion in zebrafish [17–19]. We further dissected the modulatory role of CSF-cNs onto the slow CPG during specific network states. We confirmed that activating CSF-cNs induced delayed slow locomotor activity at rest. In contrast, the activation of CSF-cNs during ongoing locomotor activity inhibited slow locomotion, reducing the duration and the frequency of locomotor events. Moreover, selective activation of rostral CSF-cNs during ongoing activity led to the silencing of activity along the full length of spinal cord, indicating that CSF-cNs primarily inhibit descending interneurons. Altogether, our results demonstrate that a single type of conserved GABAergic sensory neuron can tune the excitability of the locomotor CPG in a state-dependent manner, by modulating key excitatory premotor interneurons.

**Results**

**CSF-cNs are local, ipsilateral, and ascending GABAergic sensory neurons that innervate the ventrolateral spinal cord**

CSF-cNs have been recently shown to selectively express the transient receptor potential channel (TRP) polycystic kidney disease 2-like 1 (PKD2L1) [20–22]. To analyze the functional morphology underlying the connectivity of CSF-cNs, we cloned the *pkd2l1* promoter and generated a specific Gal4 line (Figure 1). Fluorescent *in situ* hybridization against *pkd2l1* revealed that the *Tg(pkd2l1:gal4)* line recapitulates the endogenous *pkd2l1* expression profile at 3 dpf (Figure S1). In *Tg(pkd2l1:gal4;UAS:ChR2-mCh)* double transgenic larvae, we obtained selective expression of Channelrhodopsin-2 (ChR2) in CSF-cNs throughout the entire spinal cord (Figure 1A1). CSF-cNs are characterized by an apical villiated extension contacting the central canal and an elongated soma (Figure 1A2, 1C) [23]. We found that these cells had rostrally-directed projections chiefly restricted to the ventrolateral spinal cord (Figure 1A2-A3, 1C, 1D and 1F). We confirmed the GABAergic nature of *pkd2l1*$^+$ CSF-cNs [12, 13, 21, 24] by quantifying the overlap between glutamic acid decarboxylase (GAD) expression and PKD2L1 in *Tg(pkd2l1:gal4;UAS:ChR2-mCh;gad1b:GFP)* larvae (Figure 1B, 99% of mCherry[+] CSF-cNs were GFP[+], n = 504 cells in 5 larvae).

We used sparse genetic labeling (see Supplemental Experimental Procedures) to determine the precise projection patterns of CSF-cNs axons. At 3 dpf, all CSF-cNs axons were ascending, ipsilateral, and produced local projections reaching from two to six segments away from the cell body (Figure 1C-E, n = 88 cells in 62 larvae, mean segment projection length = 3.8 ± 0.8 segments). Regarding the extent of projections within the dorsoventral (D-V) axis, we observed that the axons of CSF-cNs mainly ran in the ventral spinal cord (Figure 1C, 1F, mean D-V axon position = 0.32 ± 0.09 where the ventral limit 0 and the dorsal limit is 1). To map synaptic sites along the CSF-cN axons, we drove expression of synaptophysin-GFP to visualize putative presynaptic boutons [25]. Boutons were identified as large and stable GFP[+] clusters, while small synaptophysin-GFP-containing vesicles were dim and highly mobile (Figure 1G, Movie S1). The majority of putative synaptic sites was confined in a ventral domain between 0.2 and 0.4 on the D-V axis (Figure 1I, 60% of boutons, see the black portion of the distribution) although presynaptic boutons were observed throughout the entire CSF-cNs D-V axonal projection domain, (0-0.6 on the D-V axis; Figure 1H-I, n = 1566 boutons from 36 cells in 27 larvae, mean bouton D-V position = 0.293 ± 0.128). This observation suggests that the primary targets of CSF-cN axons lie within the ventral spinal cord.

**CSF-cNs form GABAergic synapses onto premotor V0-v glutamatergic interneurons**

The cell bodies of a subset of glutamatergic V0-v interneurons referred to as Multipolar Commissural Descending interneurons (MCoDs) in larval zebrafish precisely reside in the ventrolateral spinal cord, confined between 0.2 and 0.4 on the D-V axis, [17, 19, 26, 27]. To test whether CSF-cNs project onto glutamatergic V0-v interneurons, we took advantage of the *Tg(vglut2a:lox:DsRed:lox:GFP)* transgenic line labeling most of the glutamatergic neurons in the spinal cord [28, 29]. In this line, we identified the large soma of DsRed$^+$ ventrolateral excitatory interneurons at the interface of the lateral neuropil and the cluster of spinal interneurons (Figure 2A1-A2). In *Tg(pkd2l1:gal4;UAS:ChR2-YFP; vglut2a:lox:DsRed:lox:GFP)* larvae, axonal projections of CSF-cNs formed numerous boutons apposed onto the somata of *vglut2a$^+$* ventrolateral interneurons (top and bottom panels in Figure 2A1-A2, 2B), suggesting that CSF-cNs synapse onto these cells. Moreover, we observed that single CSF-cN axons could project onto multiple ventrolateral *vglut2a$^+$* interneurons (Figure 2B, arrowheads).

We performed targeted whole cell recordings to measure the electrophysiological properties of ventrolateral *vglut2a$^+$* interneurons with Alexa 647 dye in the recording pipette to reconstruct their morphology (Figure 2C). The morphology and firing patterns of excitatory ventrolateral *vglut2a$^+$* interneurons were similar to those described previously for glutamatergic V0-v interneurons known as MCoDs in zebrafish (Figure 2D-G) [17, 26, 27, 30, 31]. Post-hoc reconstructions of dye-filled cells demonstrated two dendrites symmetrically located on each side of the soma (Figure 2D-E, n =12 cells) and multiple dendritic ramifications (Figure 2D-E). Analysis of presynaptic boutons made by *pkd2l1*-expressing cells onto dye-filled V0-v interneurons in *Tg(pkd2l1:Gal4;UAS:ChR2-YFP)* larvae revealed dual innervation of soma and dendrites by the axons of CSF-cNs (Figure 2D, arrowheads).

We made simultaneous whole-cell current clamp recordings of $vglut2a^+$ V0-v interneurons while monitoring pooled motor output from ventral nerve root (VNR) recordings made from nearby body segments. These dual recordings showed that V0-v cells are rhythmically active during episodes of spontaneous fictive slow locomotion (15-30 Hz, Figure 2F, n = 5). In zebrafish, interneurons active during slow swimming typically exhibit input resistance greater than 400 MΩ [19]. We found that $vglut2a^+$ V0-v cells exhibited high input resistance (641 ± 52 MΩ, n = 8 cells) and that action potentials could be elicited with small depolarizing currents (Figure 2G, mean threshold current = 15.4 ± 8.3 pA, n = 12 cells) suggesting that these cells are highly excitable and recruited with relatively small levels of excitatory drive [17, 19].

To test whether CSF-cNs formed functional monosynaptic connections onto $vglut2a^+$ V0-v interneurons, we elicited single spikes in ChR2-expressing CSF-cNs using brief optical activation while recording inhibitory currents in nearby $vglut2a^+$ V0-v interneurons (Figure 3A1). Pulses of 5 ms blue (~ 460 nm) light reliably elicited single spikes in CSF-cNs (Figure 3A2, spike delay = 4.86 ± 0.50 ms, n = 141 stimulations in 4 cells). Following single light pulses, we observed GABAergic-mediated inhibitory postsynaptic currents (IPSCs) in V0-v interneurons after a short delay (Figure 3B, B1, 3C; IPSC delay = 4.94 ± 2.02 ms, time-to-peak = 1.58 ± 0.65 ms, n = 8 cells) and a decay time consistent with the de-activation of $GABA_A$ receptors (IPSC time decay = 25 ± 10.2 ms, [32]). Intriguingly, we found that the probability of observing an IPSC in response to each light pulse was initially low (Figure 3C, response probability = 0.18 ± 0.04, n = 8 cells), but increased during 500 ms train stimulations at 25 Hz to reach up to 0.5 (Figures 3B2, 3E, n = 5 cells). The probability of eliciting light-evoked IPSCs increased both during a train (Figure 3D, 3E; $p < 0.001$, n = 5 cells) and across trains over the time course of the experiment (Figure 3D, 3F, $p < 0.001$, n = 5 cells), suggesting that individual CSF-cN presynaptic terminals may have a low release probability that is overcome with repeated activation. ChR2-induced IPSCs were blocked upon bath application of gabazine (Figure 3G, n = 3 cells), indicating that CSF-cNs form $GABA_A$-mediated synapses onto V0-v interneurons. Together, these results indicate that CSF-cNs form active GABAergic synapses onto glutamatergic descending V0-v interneurons that are gradually recruited with repetitive stimulations.

**CSF-cNs exert a state-dependent modulation of the slow locomotor CPG**
Knowing that CSF-cNs project onto premotor excitatory interneurons specifically active during slow locomotion, we tested the effects of activating CSF-cNs on locomotor activity. We carefully restrained illumination to the spinal cord (see Experimental Procedures) and used a long light pulse (500 ms, Figure 4A-B) to elicit burst spiking in ChR2-expressing CSF-cNs (response delay = 7.9 ± 5.8 ms, 11.9 ± 4.3 spikes per light pulse, firing frequency = 23.8 ± 8.7 Hz, burst duration = 491.5 ± 10.6 ms, n = 2 cells). In this configuration, blue light stimulation did not trigger locomotor activity in $ChR2^-$ control siblings (Figure 4C top trace; Figure 4D, response rate = 3.6 ± 1.6 %; n = 586 stimulations in 10 larvae). In contrast, there was a significantly higher response rate following the light pulse in $ChR2^+$ larvae (Figure 4C, middle trace, 4D; response rate = 19.96 %, n = 759 stimulations in 14 larvae, $p < 0.05$). Among the fourteen $ChR2^+$ larvae tested, we observed that blue light pulses did not induce locomotor activity in six animals (Figure 4C, bottom trace, response rate 1.8 ± 0.8 %). In eight out of fourteen ChR2$^+$ larvae, the activation of CSF-cNs reliably triggered slow fictive swimming after a delay of 465 ± 55 ms (Figure 4C, middle trace, response rate = 33.6 ± 8.4 %). Such a long delay suggested that locomotor responses followed an initial period of inhibition. We tested whether the induced locomotor response was GABA-mediated on a larva with a high baseline response rate. Bath application of the GABA$_A$ receptor antagonist gabazine led to a reduction of the response rate from 0.79 to 0.16 suggesting that the rebound swimming could rely on the activation of GABA$_A$ receptors in this animal (Figure 4E).

We hypothesized that the heterogeneity of the responses observed across ChR2$^+$ larvae could be due to variations in the intrinsic excitability of the spinal locomotor circuit across fish. The excitability of spinal circuits can be modulated with the application of excitatory neurotransmitters such as NMDA [33, 34]. Bath application of low concentrations of NMDA (10-20 μM) had no effect on swimming in ChR2$^-$ control siblings (Figure 4G, left plot, response rate = 3.6 ± 1.6 % at rest and 2.4 ± 1.1 % with NMDA, n = 8 larvae). In contrast, the presence of NMDA led to a dramatic increase in the response rate to photostimulation in ChR2$^+$ larvae, by converting larvae with no response into responsive larvae (Figure 4F, compare top and bottom traces; Figure 4G, right plot, the response rate went from 6.4 ± 2.4 % at rest to 47 ± 11 % under NMDA, n = 6 larvae, p < 0.05).

We tested whether *vglut2a*$^+$ V0-v interneurons contributed to the delayed motor activity following the activation of CSF-cNs at rest. We recorded V0-v in cell-attached mode and analyzed their firing in response to blue light (Figure 4H). We found that activating CSF-cNs could induce firing in V0-v (response delay = 320 ms ± 167 ms, n= 118 out of 311 stimulations in 3 cells). Altogether these data indicate that the induction of slow locomotion at rest following CSF-cNs activation involves the recruitment of V0-v interneurons.

Because CSF-cNs form active GABAergic synapses onto glutamatergic V0-v premotor interneurons, we hypothesized that CSF-cN activation during ongoing locomotion might silence motor activity. We used a closed-loop ChR2 activation assay (Figure 5A-B) where a 500 ms light pulse was delivered at the onset of spontaneous slow fictive swimming events. In ChR2$^-$ control siblings, blue light pulses had no effect on ongoing locomotor activity (Figure 5C). In contrast, activation of ChR2$^+$ CSF-cNs led to an abrupt silencing of ongoing swimming activity (Figure 5D). The activation of CSF-cNs significantly reduced the duration of spontaneous swim bouts (Figure 5E-G, for control fish, bout duration with LED off (BD$_{LED\ OFF}$) = 550 ± 53 ms, BD$_{LED\ ON}$ = 486 ± 59 ms, n = 15 larvae; whereas for ChR2-expressing fish BD$_{LED\ OFF}$ = 840 ± 82 ms versus BD$_{LED\ ON}$ = 235 ± 33 ms, p < 0.001, n = 29 larvae). In addition, CSF-cN activation led to a significant reduction in the frequency of swim bouts (Figure 5H, for control fish, bout frequency with LED off (BF$_{LED\ OFF}$) = 0.5, BF$_{LED\ ON}$ = 0.56, n = 7 larvae; for ChR2-expressing fish, BF$_{LED\ OFF}$ = 0.52, BF$_{LED\ ON}$ = 0.33, p < 0.05, n = 6 larvae) without altering the burst frequency (bf) within bouts (Figure S2, bf$_{LED\ OFF}$ = 22.60 ± 0.64 Hz, bf$_{LED\ ON}$ 22.54 ± 0.80 Hz). These findings demonstrate a marked effect of CSF-cN activation on the duration and occurrence of spontaneous bouts of fictive swimming.

We tested whether the silencing of locomotor activity was mediated by the activation of ionotropic $GABA_A$ receptors. Bath application of $GABA_A$ blocker gabazine (10-20 µM) led to an increase in the spontaneous bout duration (compare the left panel of Figure 5I with the left panel of Figure 5J; Figure 5K, $BD_{LED\ OFF}$ = 909 ms ± 117 ms, $BD_{LED\ OFF/gabazine}$ = 2.18 s ± 0.64 s, p < 0.0001, n = 10 larvae) indicating that the endogenous release of GABA modulates the duration of locomotor events. Nonetheless the presence of gabazine reduced the silencing mediated by CSF-cNs (compare Figure 5I right panel with Figure 5J right panel, Figure 5K, $BD_{LED\ ON}$ = 184 ± 38 ms, $BD_{LED\ ON/gabazine}$ = 779 ± 128 ms, p < 0.0001). The silencing of locomotion by CSF-cNs being independent of the initial bout duration (Figure S3), we measured the silencing efficiency of CSF-cNs, i.e. their ability to reduce bout duration, as the ratio of bout duration with or without the activation of $ChR2^+$ CSF-cNs ($BD_{LED\ OFF}/BD_{LED\ ON}$). The silencing efficiency was reduced from 6.9 to 2.9 when gabazine was added into the bath (Figure 5L, p < 0.05, n = 10), indicating that the inhibition of fictive locomotion involves the activation of $GABA_A$ receptors.

Together, these results demonstrate that CSF-cNs can modulate the slow locomotor CPG in a state-dependent manner. CSF-cNs can trigger delayed rhythmic activity at rest, an effect enhanced by 10-20 µM NMDA in the bath. In contrast, CSF-cNs can inhibit ongoing locomotor activity, reducing both the duration and the frequency of occurrence of locomotor events.

**Inhibition of locomotor activity by CSF-cNs predominantly occurs at the premotor level**

The suppression of ongoing locomotor activity could be due to the direct inhibition of motor neurons or/and of descending premotor excitatory interneurons controlling slow locomotion. We first tested whether CSF-cN activation could silence ongoing activity in V0-v excitatory interneurons recorded in cell-attached mode (Figure 6A). We observed that spontaneous firing events in V0-v interneurons were silenced by CSF-cN activation (Figure 6B) leading to shorter episodes (Figure 6C-D; mean event duration with LED off ($ED_{LED\ OFF}$) = 341 ± 50 ms, $ED_{LED\ ON}$ = 176 ± 77 ms, n = 99 and 77 episodes respectively, recorded in n = 3 cells) containing fewer spikes (Figure 6E, mean spikes per episodes with $LED_{OFF}$= 15.8 ± 4.9 spikes, 7.7 ± 2.8 spikes/episodes with $LED_{ON}$). These data reveal that CSF-cNs can silence locomotor activity at the premotor level.

Since glutamatergic V0-v interneurons project 15-20 segments caudally [17, 27] and are silenced by CSF-cNs, we hypothesized that silencing rostral V0-v interneurons could alter the propagation of locomotor activity to more caudal motor pools. To test this hypothesis, we restricted the optical activation of CSF-cNs to rostral (1-10) or caudal (16-25) segments while performing dual VNR recordings at rostral (5-8) and caudal (15-16) segments (Figure 7A-C). The activation of rostral CSF-cNs strongly silenced fictive swimming activity in rostral as well as in caudal segments, occasionally abolishing caudal motor activity (Figure 7B, see stars on bottom traces, 7D, n = 6 larvae). In contrast, the activation of caudal CSF-cNs had small effects on both rostral and caudal motor activity compared to the activation of rostral CSF-cNs (Figure 7C, 7D, n = 4 larvae, Figure 7E silencing efficiency for rostral stimulations = 2.85 at VR 8 and 2.04 at VR 16, silencing efficiency for caudal stimulations = 1.08 at VR 8 and 1.06 at VR 16, p < 0.01). The activation of rostral CSF-cNs did not seem to modulate the rostrocaudal lag measured between segment 8 and 16 (Figure S4).

Altogether, these results suggest that the ability of CSF-cNs to disrupt the excitatory drive along the spinal cord most likely rely on the modulation of premotor interneurons as these cells have long descending projections within the spinal cord.

**Discussion**

In the present study, we investigated the cellular and network mechanisms underlying the modulation of slow locomotion by cerebrospinal fluid-contacting neurons in larval zebrafish. To our knowledge, this work is the first investigation of the functional connectivity of a single GABAergic neuron type onto glutamatergic interneurons of the locomotor CPG. Our results highlight the complexity of this GABAergic modulatory pathway leading to antagonistic effects depending on the excitability or state of spinal motor circuits.

**Circuit organization of spinal cerebrospinal fluid-contacting neurons: projections onto excitatory elements of the slow swimming CPG**

Taking advantage of existing transgenic lines labeling glutamatergic interneurons [28, 29], we observed anatomical and functional connections from CSF-cNs onto *vglut2a*$^+$ V0-v premotor interneurons in the spinal cord. In larval zebrafish, these cells are specifically active during slow locomotion with bursting frequencies ranging from 15 to 30 Hz and are selectively silenced at swimming frequencies above 30 Hz, possibly by glycinergic interneurons [17]. Our results demonstrate that additional GABAergic inputs to *vglut2a*$^+$ V0-v originate from CSF-cNs. Given the increasing fidelity of ChR2-driven IPSCs between CSF-cNs and V0-v during 25 Hz train stimulations,

CSF-cNs may also contribute to the frequency-dependent suppression of V0-v activity. In mice, although glutamatergic V0-v interneurons are essential components of the locomotor CPG [15, 16], they appear critical for left-right alternation during fast locomotion [35], and their ablation selectively affects trot [36]. Further work will be necessary in zebrafish to test whether CSF-cNs can modulate left-right alternation.

Interestingly, activation of CSF-cNs reduced the occurrence of locomotor events. This effect could be explained by the modulation of supraspinal neurons thought to control the initiation and frequency of locomotor events [37, 38]. In the rostral spinal cord, we observed that CSF-cNs projected into the lateral margins of the caudal hindbrain, reaching the somata of V0-v interneurons as well as descending excitatory fibers projecting in the spinal cord. CSF-cNs projecting onto the caudal hindbrain could delay the occurrence of locomotor events by silencing the output of hindbrain interneurons projecting in the spinal cord. However, the full connectivity map of CSF-cNs on their targets, including neurons located in the hindbrain, remains to be completed.

Regarding the neurotransmitter released by CSF-cNs, the modulation of locomotion by CSF-cNs was at least partially mediated by $GABA_A$ receptors. However, there might be additional components to the effect mediated by CSF-cNs. First, gabazine in the bath failed to fully abolish the silencing of ongoing locomotion. Second, the relatively short inactivation time of $GABA_A$ receptors (<100 ms) does not match the long lasting effects (seconds) of increasing inter bout interval. CSF -cNs have been shown to express a variety of peptides [24, 39]. It is therefore plausible that other receptors for GABA and/or peptides also contribute to the modulation of locomotor activity by CSF-cNs.

**State-dependent modulation of locomotor activity**

Using electrophysiology and pharmacology, our study sheds light on the state-dependent GABAergic modulation of locomotor CPGs by CSF-cNs. On one side, we revealed the inhibitory action of CSF-cNs when stimulated during ongoing locomotion. In this context, locomotor activity was silenced within 200 ms on average, suggesting that a build up of inhibition was necessary. On the other side, activation of CSF-cNs at rest induced delayed fictive swimming that was highly dependent on the intrinsic excitability of the spinal cord. The delay of induced swimming was about 450 ms, ruling out a direct activation of locomotor CPGs. One possible explanation is that rebound activity originates from an accumulation of depolarizing inhibition as depolarizing GABA is common in immature spinal circuits [40]. Alternatively, the induction of swimming may follow a rebound from $GABA_A$-mediated inhibition. Post-inhibitory rebound (PIR), a general feature of rhythmic networks including locomotor CPGs [41–43], has been proposed to regulate the timing of activation of premotor interneurons [41, 42, 44–46]. In the tadpole, PIR is an emergent property of a complex interplay of inhibition and depolarization that is modulated in a state-dependent manner [43]. Even though our data indicate that V0-v contribute to the delayed activity triggered by CSF-cNs at rest, it is unlikely that PIR originates solely from the intrinsic properties of these interneurons as we did not observe post-hyperpolarization rebound spiking in these cells. Following CSF-cN activation, PIR may rise from network interactions via other targets, leading to the rebound firing of V0-v interneurons and subsequent induction of slow swimming.

**Roles for a CSF-dependent GABAergic inhibition of premotor excitation**

Since our experiments relied on forcing the activation of CSF-cNs with light, one open question lies in identifying the physiological conditions and the timing under which these cells are normally recruited *in vivo*. Kolmer initially thought that CSF-cNs could form a parasagittal organ functioning as a third ear in the spinal cord [9]. Indeed, the morphology of these cells extending in the central canal is optimal to detect chemical or mechanical cues from the CSF. Previous studies in mammals indicated that CSF-cN firing was modulated by changes of extracellular pH [20, 47] but how such information is transduced and relates to locomotion is still unclear. The observation that the reliability of CSF-cNs to V0-v synaptic currents increases during 25Hz train implies that the silencing mediated by CSF-cNs is particularly efficient with persistent spiking in the range of slow swimming frequencies. The inhibition of CSF-cNs onto V0-v interneurons could thereby build up over time during locomotor events. Future work will be necessary to complete the connectivity pattern and modulatory role of CSF-cNs in fish and mammals in order to elucidate CSF-cN modulatory function of active locomotion in vertebrates.

**Experimental Procedures**

**Animal care**

Animal handling and procedures were validated by ICM and the National Ethics Committee (*Comité National de Réflexion Ethique sur l'Expérimentation Animale*-Ce5/2011/056) in agreement with E.U legislation. Adults were reared at a maximal density of 8 animals per liter in a 14/10 hours light/dark cycle environment. Fish were fed live artemia twice a day and feeding regime was supplemented with solid extracts matching the fish developmental stage (ZM Systems, UK). Larvae were raised at 28.5°C with a 14/10 day/night light cycle. Experiments were performed at room temperature (22-25°C) on 3 to 5 dpf larvae.

**Generation of transgenic animals**

Transgenic lines used in this study are listed in Table S1. Detailed procedures including the generation of transgenic animals, FISH combined with IHC and the morphological analysis of CSF-cNs are available in Supplemental Experimental Information.

**Live imaging of spinal neurons**

Zebrafish larvae were imaged using an upright microscope (Examiner Z1, Zeiss, Germany) equipped with a spinning disk head (CSU-X1, Yokogawa, Japan) and a modular laser light source (LasterStack, 3i Intelligent Imaging Innovations, Inc., Denver, Colorado, USA). Z-projections stacks were acquired using Slidebook software (3i) and reconstructed online using Fiji (http://fiji.sc/Fiji).

**Sample preparation for fictive locomotion recordings and optogenetic stimulations**

3 dpf *Tg(pkd2l1:gal4;UAS:ChR2-YFP)* larvae were screened for dense labeling and bright expression of ChR2-YFP in CSF-cNs under a dissecting fluoroscope (Leica, Germany). Larvae were anaesthetized in 0.02% Tricaine-Methiodide (MS-222, Sigma-Aldrich, USA) diluted in fish facility water and then mounted upside-down (ventral side facing up) in glass-bottom dishes (MatTek, Ashland, Massachusetts,

USA) filled with 1.5 % low-melting point agarose. We surgically removed the eyes using a thin tunsgten pin in order to avoid light-evoked locomotion with blue light during ChR2 stimulation. Following the surgery, larvae were transferred in cold ACSF (concentrations in mM: 134 NaCl, 2.9 KCl, 1.2 $MgCl_2$, 10 HEPES, 10 glucose and 2.1 $CaCl_2$; 290 mOsm, adjusted to pH 7.7-7.8 with NaOH) for 3-5min. Larvae were then transferred in fish facility water to recover for 24 hours at 28°C. The following day, larvae were mounted on their side and immobilized by injecting 0.5 nl of 0.5 mM α-Bungarotoxin in the ventral axial musculature (Tocris, Bristol, UK). A portion of agarose was removed using a sharp razorblade in order to expose 2 to 3 segments.

**Fictive locomotion recordings and optogenetic stimulation**

Our recording protocol is based on published procedures [17, 19, 48]. VNR recordings were acquired using a MultiClamp 700A amplifier, a Digidata series 1322A Digitizer and pClamp 8.2 software (Axon Instruments, Molecular Devices, Sunnyvale, CA, USA). A blue LED (UHP-Mic-LED-460, Prizmatix Ltd., Modiin-Ilite, Israel) was used to activate ChR2. The light was delivered on the fish spinal cord through the microscope condenser, typically 20 segments from segment 7/8 to 27/28 with 14 $mW/mm^2$. To time the optical activation of ChR2 after the onset of fictive swimming bouts, we designed a closed-loop program in which the LED was turned on via TTL pulses 10 ms or 500 ms after the fictive motor output reached an arbitrary threshold. Parameters describing the fictive locomotion were extracted using custom-made MATLAB scripts. Bout frequency was analyzed in larvae with basal level of swimming activity above 0.3 Hz.

**Pharmacology**

4 dpf larvae were pinned-down in Sylgard-coated, glass bottom dishes filled with ACSF with thin tungsten pins through the notochord. Skin was removed from segments 5-6 to the end of the tail using sharp forceps. $GABA_A$ receptor antagonist gabazine (SR95531, Tocris) or NMDA (Tocris) were bath applied at either 10 or 20 µM final concentrations.

**Fluorescence-guided whole-cell recordings**

Whole cell recording were performed in head-off larvae in the same configuration as pharmacology experiments. After removing the skin, one to two segments were dissected using glass suction pipettes. Patch pipettes (1B150F-4, WPI, Sarasota, FL, USA) were designed to reach a tip resistance of 11-15 MΩ and were filled with potassium-containing internal solution (concentrations in mM: K-gluconate 115, KCl 15, $MgCl_2$ 2, Mg-ATP 4, HEPES free acid 10, EGTA 0.5, 290 mOsm, adjusted to pH 7.2 with KOH and supplemented with Alexa 647 at 4 µM final concentration). To resolve evoked inhibitory postsynaptic currents in voltage-clamp mode, cells were held at around -80 mV, away from the calculated chloride reversal potential (-51mV). We calculated the liquid junction potential in our experiments (-19mV) but did not correct for it since it did not affect the outcome of our experiments. Kinetic parameters of light-evoked currents and IPSCs were extracted and analyzed using custom-made MATLAB scripts.

**Statistics**

Linear correlation in datasets was calculated using a Pearson's linear correlation test. Comparisons between two groups of data were performed using a Student's t test. A linear mixed effects model was used to test the interaction between the LED and gabazine. The level of significance was $p < 0.05$ for all datasets.

**Author Contributions**

KF performed electrophysiological recordings, pharmacology experiments, and imaging of spinal lines with the help of CS. LD and AP performed single cell morphology analysis. FDB and AB generated transgenic animals. JG performed FISH experiments. KF and CW designed experiments, analyzed data and wrote the manuscript.

**Acknowledgments**

We thank Prof. Shin-Ichi Higashijima for kindly sharing transgenic lines. We thank Prof. Kawakami for sharing reagents. We thank Dr. Jean Simonnet for helping setting up patch-clamp recordings. We thank Dr. Richard Miles, Dr. Alberto Bacci, Dr. Daniel Zytnicki, Prof. Eve Marder and members of the Wyart lab for critical reading of the manuscript. We thank Natalia Maties, Bodgan Buzurin and Sophie Nunes-Figueiredo for fish care. This work received support from ICM, Ecole des Neurosciences de Paris, Fondation Bettencourt-Schueller, Mr Pierre Belle, City of Paris, Atip/Avenir program, Marie Curie Actions (IRG #227200), and ERC starting grant Optoloco (#311673).

# Figures

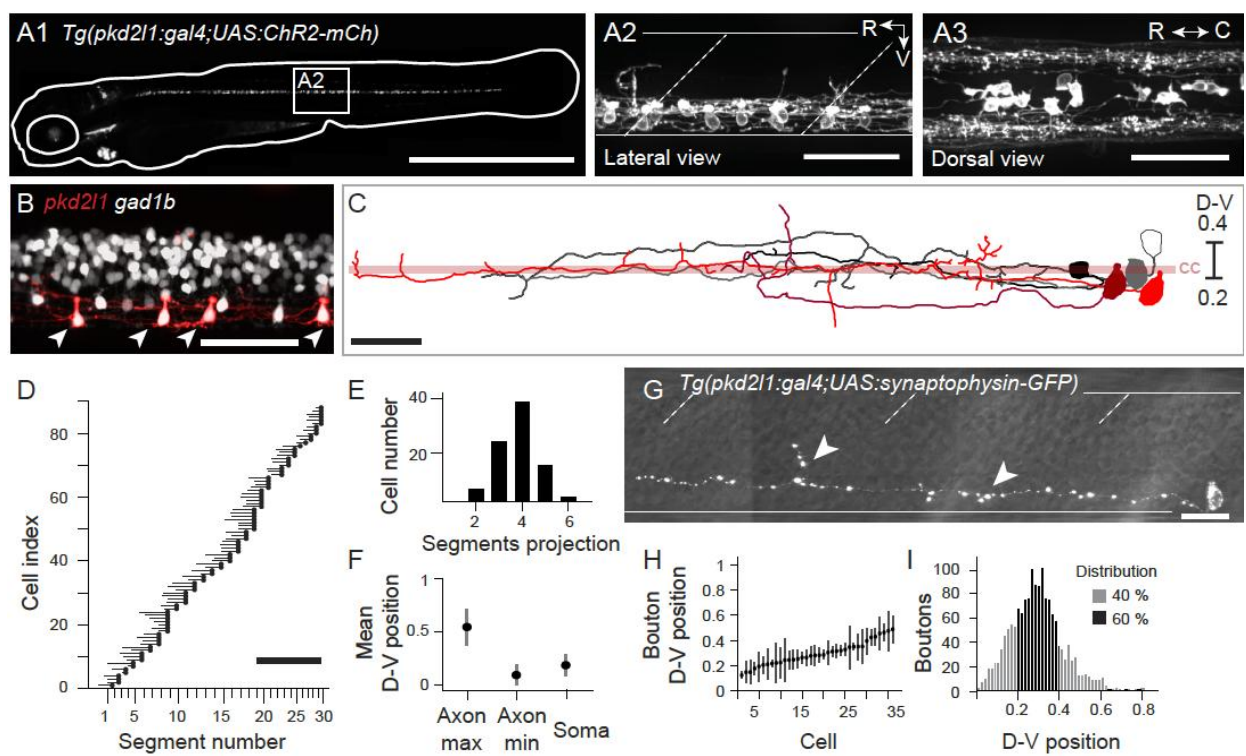

**Figure 1. The *Tg(pkd2l1:gal4)$^{icm10}$* stable transgenic line specifically labels CSF-cNs in the zebrafish spinal cord**

(A1) Complete pattern of expression of mCherry in *Tg(pkd2l1:gal4;UAS:ChR2-mCh)* double transgenic larvae at 4 dpf. (A2) Lateral view of the spinal cord shows that mCherry was restricted to CSF-cNs in the ventral part of the spinal cord. Seven CSF-cNs were labeled per axial segment on average. (A3) Dorsal view of the spinal cord shows that axonal projections of CSF-cNs were ipsilateral and located in the lateral margins of the spinal cord. See also Figure S1 and Table S1.

(B) Overlap of mCherry (red) and GFP (grey) in the *Tg(pkd2l1:gal4;UAS:ChR2-mCh;Gad1b:GFP)* triple transgenic larvae confirms the GABAergic nature of CSF-cNs (arrowheads). Note that the *Tg(pkd2l1:gal4)* line does not label all CSF-cNs.

(C) Morphology of five CSF-cNs located in segment 9 at 3 dpf after single cell imaging and reconstruction. Axonal projections varied in length, in branching as well as in dorsoventral (D-V) positioning. Cells were aligned according to the D-V position of their cell body. The central canal (cc) is represented by the light red bar.

(D) Mapping of CSF-cNs axonal projections across the rostrocaudal (R-C) axis. All CSF-cNs had ascending projections reaching from two to six segments away from the cell body (black circles, n = 88 cells).

(E) Distribution of the number of segments covered by single axons of CSF-cNs.

(F) Mean D-V positions of axons (maxima and minima) and soma of CSF-cNs.

(G) Single synaptophysin-GFP$^+$ CSF-cN with punctate synaptophysin clustering consistent with presynaptic boutons distributed along the entire axonal arborization (arrowheads). See also Movie S1.

(H) D-V position of putative synaptic boutons for CSF-cNs sorted according to their soma position along the D-V and the R-C axes (n = 36 cells).

(I) Distribution of putative synaptic boutons along the D-V axis (black bars indicate that 60 % of the boutons are confined in the 0.2-0.4 interval).

In (A2, G) white solid lines delineate the ventral and dorsal limits of the spinal cord, dashed lines for the limits of axial segments. Scale bars are 1 mm in (A1), 50 µm in (A2, A3, B), 900 µm in (D), 20 µm in (C, G). R is rostral, C is caudal, V is ventral, CC is central canal. (A-B) were reconstructed from Z-stack projections through the entire spinal cord. Data are represented here as mean ± SD.

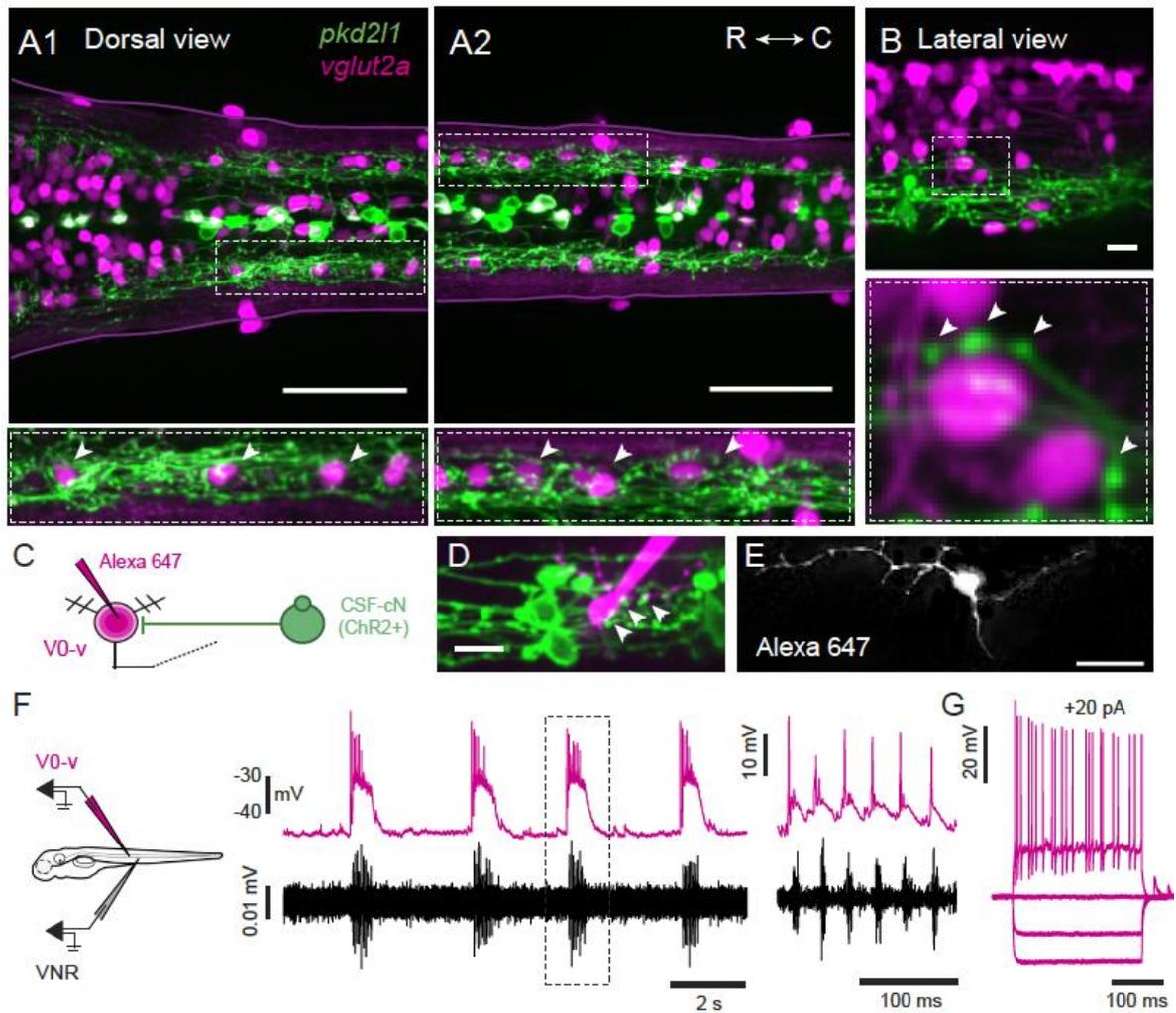

**Figure 2. CSF-cNs project onto V0-v glutamatergic interneurons**

(A1, A2) Z-stack projections of a few optical sections imaged from the dorsal side in *Tg*(*pkd2l1:gal4;UAS:ChR2-YFP; vglut2a:lox-DsRed-lox-GFP*) triple transgenic larvae reveal the localization of glutamatergic neurons (DsRed[+], in magenta) relative to CSF-cNs projections (YFP[+], in green) in rostral (A1) and caudal spinal cord (A2). Zoom of lateral regions circled in dashed lines show that CSF-cNs projections surround the DsRed[+] nucleus of ventrolateral glutamatergic interneurons (arrowheads in bottom panels (A1-A2)). See also Table S1.

(B) Apposition of CSF-cNs axonal varicosities onto the cell bodies of two ventrolateral *vglut2a*[+] interneurons (arrowheads).

(C) Ventrolateral *vglut2a*$^+$ V0-v interneuron receiving projections from CSF-cNs were filled with Alexa 647 in order to image and reconstruct their morphology.

(D) Z-stack projection of a few optical sections imaged from the lateral side in a *Tg(pkd2l1:gal4;UAS:ChR2-YFP;vglut2a:lox-DsRed-lox-GFP)* triple transgenic larva labeling CSF-cNs (ChR2-YFP$^+$, in green) and glutamatergic interneurons (DsRed$^+$, in magenta) after dye filling a *vglut2a*$^+$ V0-v interneuron. Arrowheads highlight axonal projections of CSF-cNs onto the soma and dendrites of the filled V0-v interneuron.

(E) Typical morphology of ventrolateral *vglut2a*$^+$ V0-v interneurons filled with Alexa 647.

(F) Paired ventral nerve root recording (VNR) with whole cell current clamp recording of a V0-v interneuron showing rhythmic activity during every episode of fictive slow locomotion (burst frequency ranged between 15 and 30 Hz). The region circled with dashed line is zoomed on the right to emphasize the activity of V0-v during a fictive bout. V0-v action potentials preceded motor neurons spiking for all locomotor bursts.

(G) V0-v cells are electronically compact, typically requiring only 10-20 pA of current injection to reach AP-firing threshold. Note that the cell depicted in G is not the same cell as in F.

Scale bars are 50 µm in (A1, A2) and 10 µm in (B, C and E). In each panel, Z-stack projections were reconstructed from a few optical sections (depth = 0.55 µm).

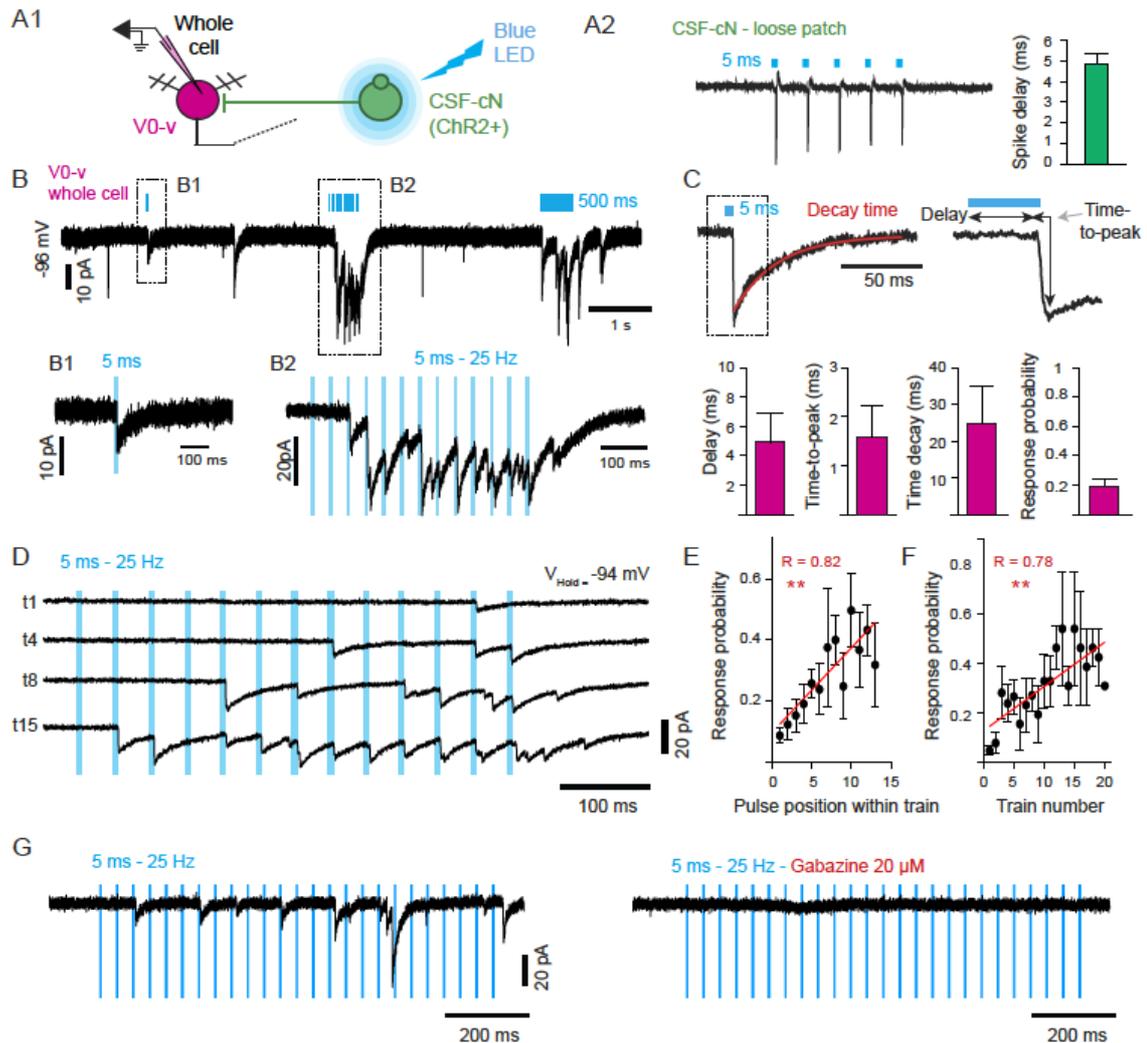

**Figure 3. CSF-cN form active GABAergic synapses onto V0-v glutamatergic interneurons**

(A1) Experimental paradigm. Targeted whole cell voltage-clamp recordings from *vglut2a+* V0-v interneurons while activating ChR2[+] CSF-cNs using brief (5ms) pulses of blue light. (A2) Loose-patch recording of a ChR2[+] CSF-cN shows that 5 ms blue light pulses (blue bars) reliably triggered single action potentials in CSF-cNs. The plot indicates that the spike delay relative to the onset of the light pulse was about 4-5ms (141 stimulations from 4 cells).

(B) Voltage clamp recording (Vm = -96 mV; reversal potential for chloride = -51 mV, therefore IPSCs appear as inward currents) of a V0-v interneuron during optical stimulation of ChR2+ CSF-cNs (represented by the blue bars) using a single pulse and a train of 13 pulses at 25 Hz.

(C) Quantification of kinetic parameters of ChR2-induced IPSCs in V0-v interneurons: delay was calculated relative to the onset of the light pulse, decay time was obtained after fitting the decay with a single exponential fit, time-to-peak was calculated as the difference between the time of current onset and of current peak.

(D) Reliability of IPSC events increased for subsequent pulses during a trial as well as across subsequent trials recorded every 20-25 s (1st, 4th, 8th and 15th trials shown).

(E) Quantification of the response probability in V0-v interneurons for each pulse during a train.

(F) Quantification of the response probability in V0-v interneurons, averaged per pulse during each train, as a function of the train number over the time course of the experiment.

(G) ChR2-induced IPSCs in V0-v interneurons are blocked by application of gabazine in the bath (n = 3 cells). Blue bars represent the stimulation at 25 Hz.

Red lines in (E) and (F) are linear fits. **, p value < 0.001 following Pearson's linear correlation test. R is the linear correlation coefficient. Data are represented as mean ± SEM.

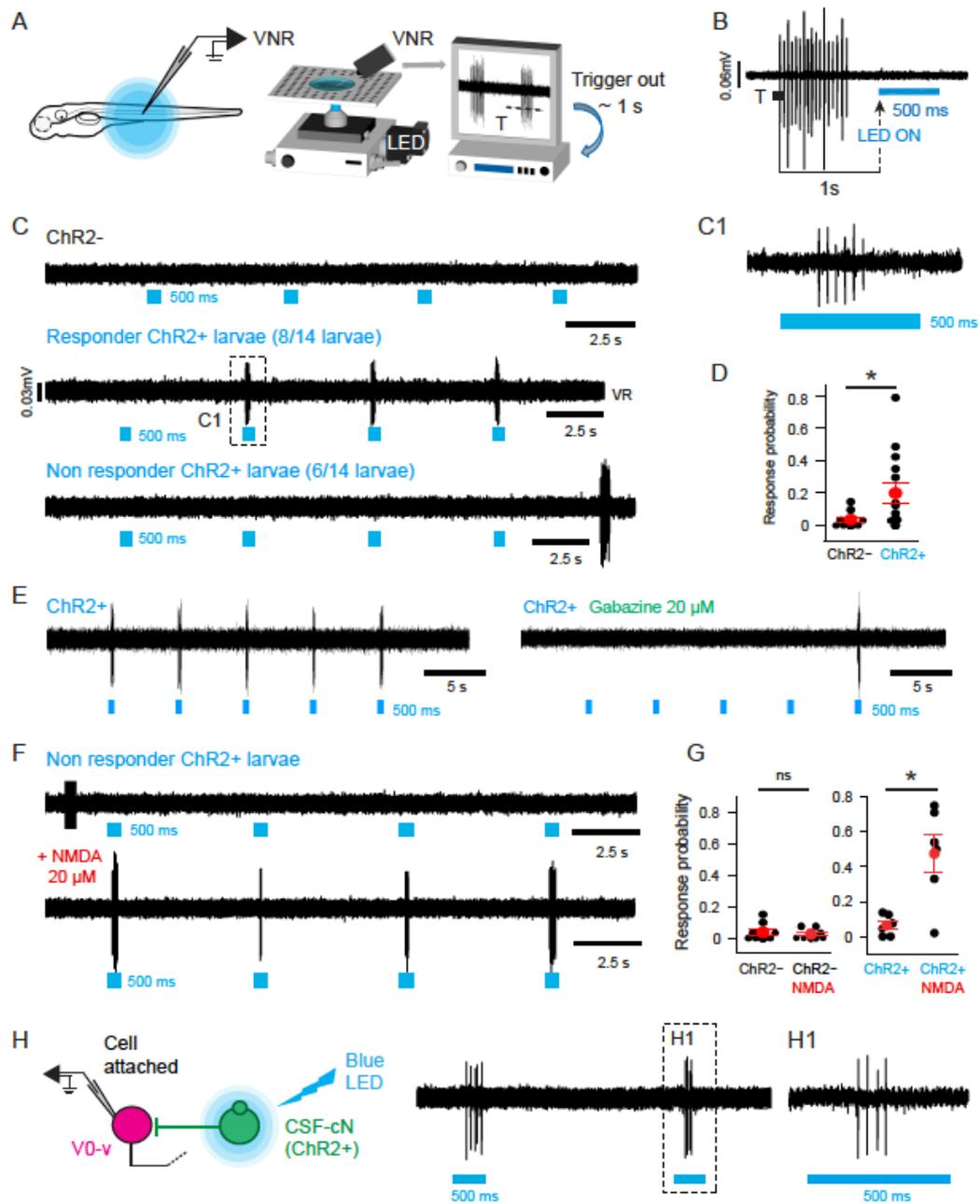

**Figure 4. Activation of CSF-cNs at rest can trigger delayed slow locomotion**

(A) Experimental paradigm showing an eye-enucleated 4 dpf ChR2$^+$ larva mounted on its side in a glass-bottom dish after paralysis. Blue light was patterned onto the spinal cord from segment 7/8 to 27/28 through the microscope condenser. VNR signals were recorded from the axial musculature and analyzed in real time. A threshold (T) was set to trigger the LED based on the VNR signal.

(B) Photoactivation protocol. LED was triggered during resting period when the animal was not fictively swimming. If the larva was not swimming, the LED was automatically activated every 4-5 seconds. If the larva was swimming, the LED was triggered long (500 ms to 2 s) after the onset of fictive bouts. The blue bar represents the pulse of blue light.

(C) Sample traces of VNR recordings showing the effects of activating ChR2$^+$ CSF-cNs using blue light at rest (light pulses are represented by the blue bars). ChR2$^-$ control siblings never showed swimming activity in response to blue light stimulations (top, n = 10 larvae). A responding ChR2$^+$ responder larva showed a delayed swimming response after the onset of the light pulse (middle, n = 8 out of 14; zoom in C1). Some ChR2$^+$ larvae did not respond to the light stimulation at rest (bottom, n = 6 out of 14).

(D) Quantification of the probability to induce delayed fictive swimming after blue light stimulations in ChR2$^-$ and ChR2$^+$ larvae (n = 10 and 14 respectively).

(E) Sample VNR traces illustrating that the induction of delayed swimming in a responsive ChR2$^+$ larva was blocked after the addition of 20 µM gabazine in the bath.

(F) Sample traces showing that 20 µM NMDA bath application transformed ChR2$^+$ non-responder larvae into responder larvae. Note that after NMDA application, the light response of the ChR2$^+$ non-responder larva mimicked the response observed in responder larvae (C, middle).

(G) Quantification of the probability to induce delayed fictive swimming responses in ChR2$^-$ and ChR2$^+$ larvae before and after NMDA application.

(H) Targeted cell attached recordings from a *vglut2a*$^+$ V0-v interneuron while activating CSF-cNs expressing channelrhodopsin-2 (ChR2) using 500 ms pulses of blue light. Delayed firing is triggered by blue light stimulations. 118 out of 311 stimulations led to delayed firing in n = 3 V0-v cells. The circled region is zoomed in panel H1 to better resolve the spiking.

*, $p < 0.05$ following Student's t test. Data are represented as mean ± SEM. Mean values are depicted in red.

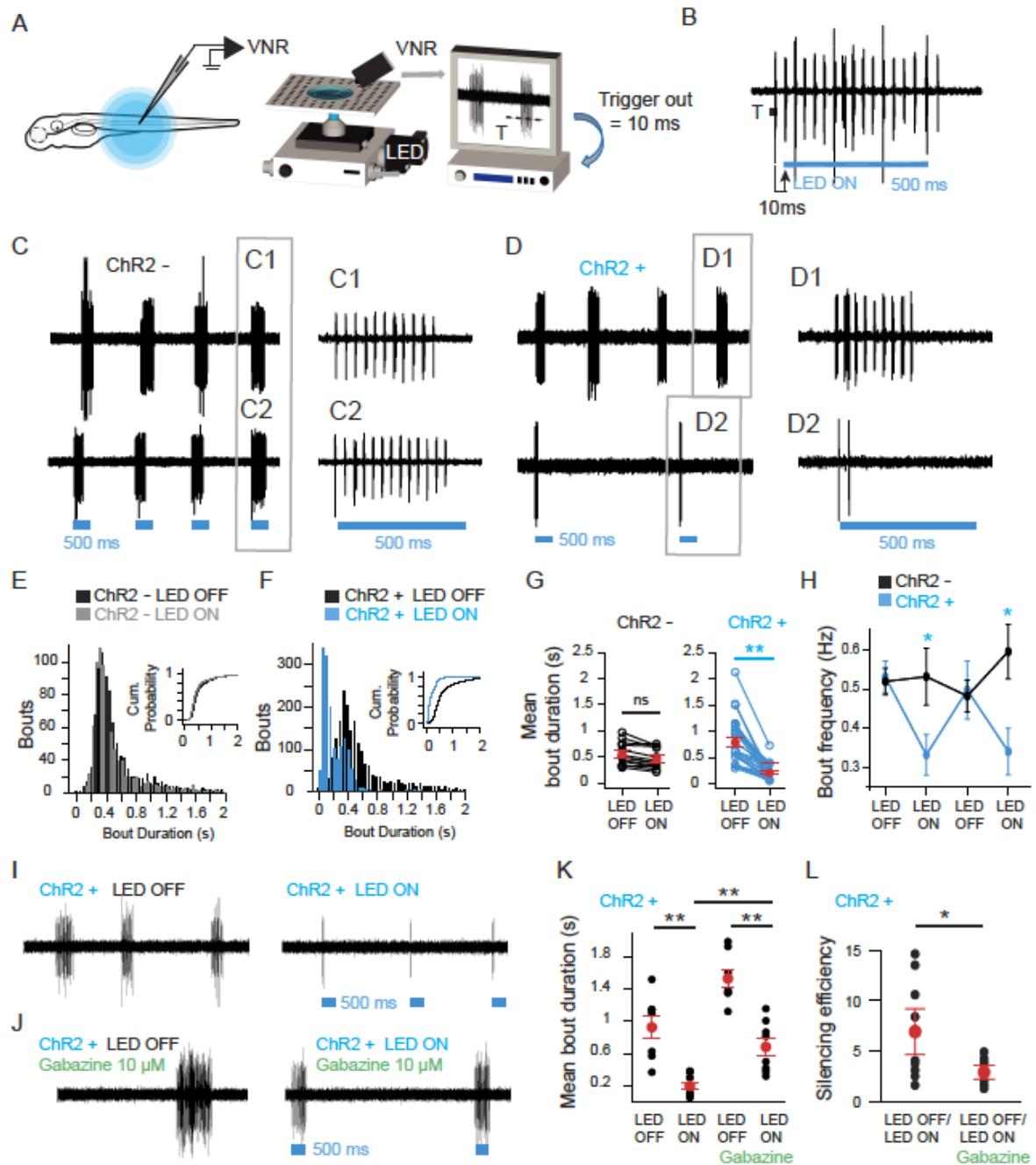

**Figure 5. Activation of CSF-cNs at the onset of ongoing fictive swimming silences locomotor activity**

(A) Experimental paradigm. Same as in Figure 4A, except that the LED was triggered at the onset of each fictive swim.

(B) Photoactivation protocol. LED was triggered rapidly (10ms) after the onset of spontaneous fictive bouts that were detected when the VNR signal had reached a manually defined threshold (T). The blue bar represents the light pulse.

(C) In ChR2⁻ control siblings, profiles of fictive locomotor activity without (top) or with (bottom) blue light stimulations (blue bars) were similar (zooms in (C1, C2)).

(D) Same as in (C) in ChR2⁺ larvae: a reduction of bout duration (BD) and bout frequency (BF) is associated with the optical activation of CSF-cNs (zoom on bouts in (D1,D2)).

(E) Distribution of spontaneous fictive bout durations in ChR2⁻ control siblings was similar with $LED_{OFF}$ (black bars) or $LED_{ON}$ (grey bars). The inset shows the cumulative probability of fictive bout duration in $LED_{OFF}$ and $LED_{ON}$ conditions.

(F) Same as in (E) in ChR2⁺ larvae. In contrast to ChR2⁻ control siblings, the distribution and cumulative probability were different in $LED_{OFF}$ (black bars) and $LED_{ON}$ (blue bars) conditions, indicating a large decrease of bout duration when the LED was ON.

(G) Quantification of the mean bout duration in both conditions for ChR2⁻ and ChR2⁺ larvae.

(H) Quantification of the bout frequency (BF) in ChR2⁻ and ChR2⁺ larvae in recordings where the LED was alternatively ON and OFF. In control ChR2⁻ larvae, the frequency did not change over the time course of the experiment whereas the bout frequency reversibly decreased with $LED_{ON}$ in ChR2⁺ larvae. See also Figure S2.

(I) Sample VNR traces in a ChR2⁺ larva illustrating the typical silencing mediated by CSF-cNs.

(J) Same as in (I) after bath application of 10 μM gabazine. Note that bout duration increased upon application. The drug penetrated the spinal cord about 5 minutes after application. Data were collected between 5 and 15 minutes after application, before seizures occurred.

(K) Quantification of fictive bout duration in ChR2$^+$ larvae during either LED$_{OFF}$ or LED$_{ON}$ episodes as well as before and after gabazine was bath-applied.

(L) CSF-cNs-mediated silencing efficiency, calculated as the relative reduction of bout duration after CSF-cNs activation (BD $_{LED\ OFF}$/BD $_{LED\ ON}$ ratio), decreased upon gabazine treatment. See also Figure S3.

In (G), (H) and (L), *, $p < 0.05$, **, $p < 0.0001$ following Student's t test. In (K), a linear mixed effects model was applied to compare datasets, and post-hoc multiple comparisons of means were performed to extract p-values (**, $p < 0.001$). Data are represented as mean ± SEM. In G, K, and L mean values are depicted in red.

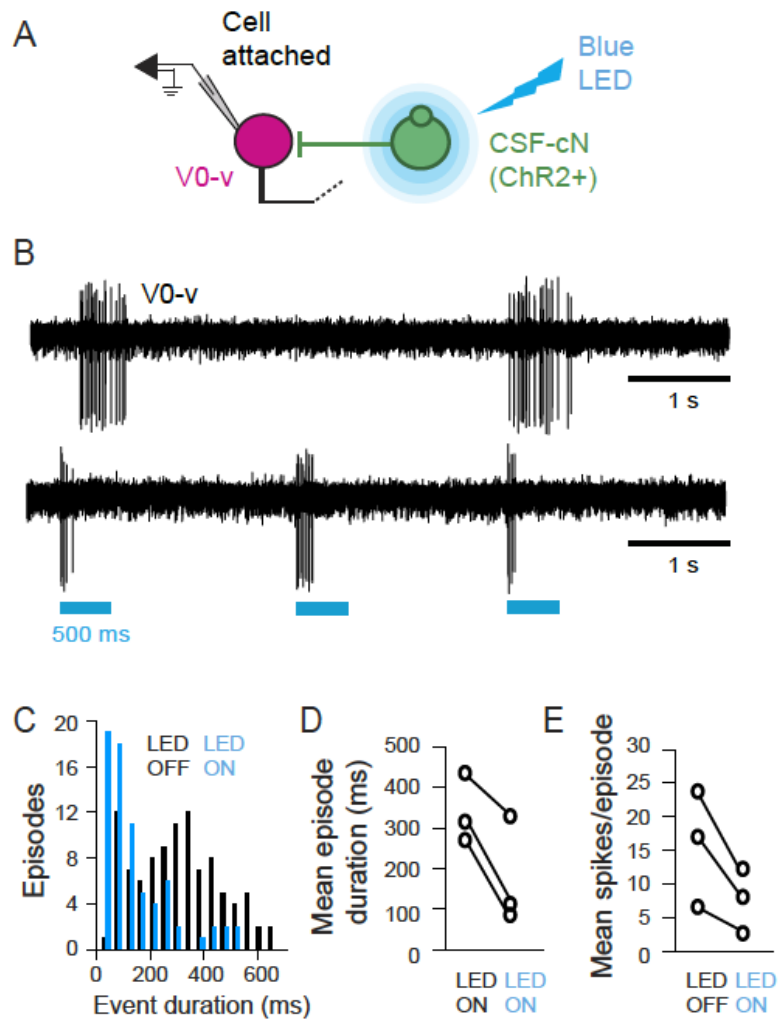

**Figure 6. Activation of CSF-cNs silences episodes of glutamatergic V0-v interneurons activity**

(A) Experimental paradigm: targeted cell attached recordings from *vglut2a$^+$* V0-v interneurons while activating CSF-cNs expressing channelrhodopsin-2 (ChR2) using 500 ms pulses of blue light.

(B) Episodes of V0-v activity (top: spontaneous activity; bottom: with stimulation of CSF-cNs timed to the onset of V0-v episodes). Blue bars represent blue light pulses.

(C) Distribution of V0-v episode durations with LED$_{OFF}$ or LED$_{ON}$.

(D) Mean episode duration per cell with LED$_{OFF}$ or LED$_{ON}$.

(E) Mean number of spikes per episodes per cell with LED$_{OFF}$ or LED$_{ON}$.

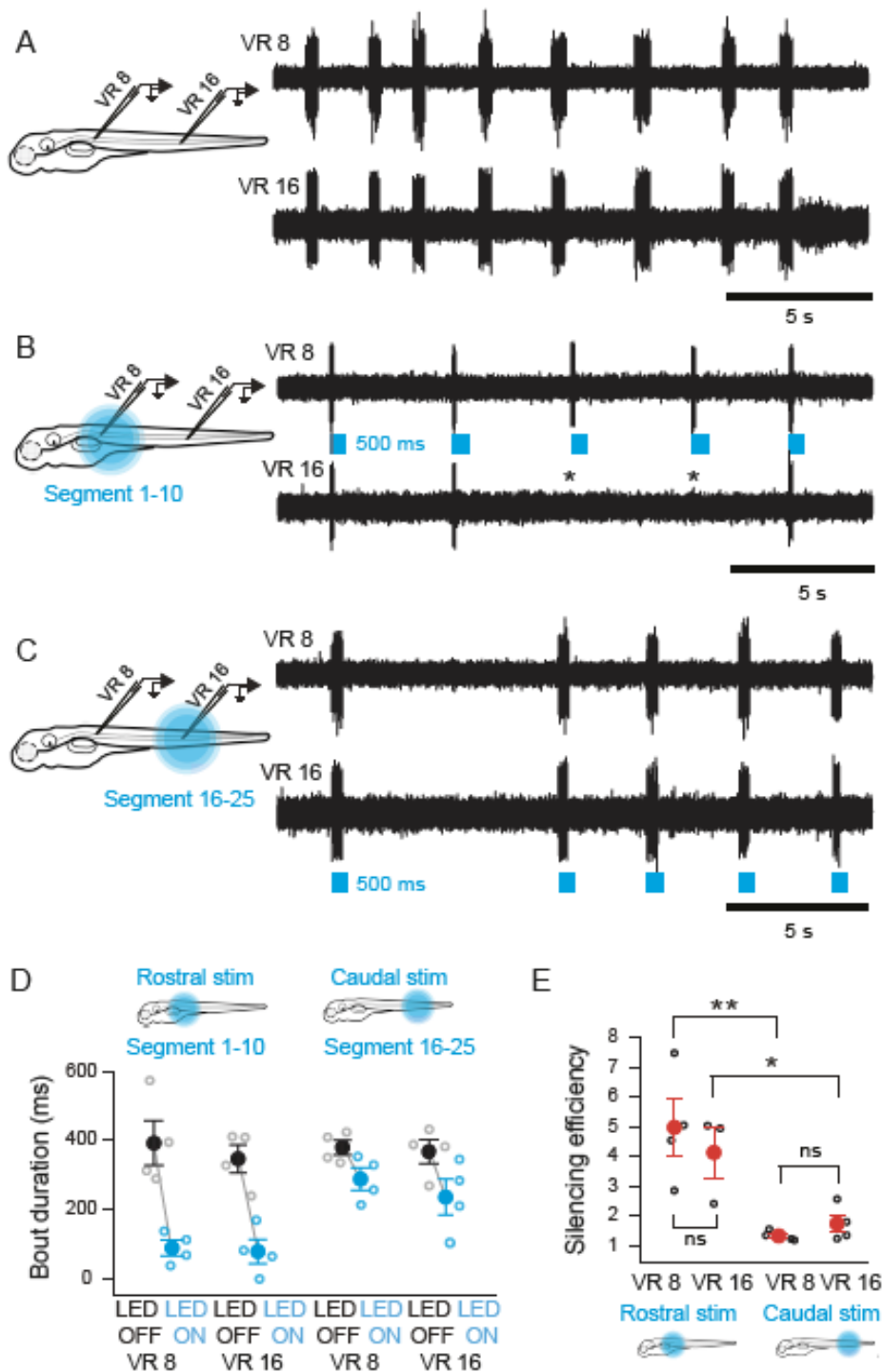

**Figure 7. Selective activation of rostral CSF-cNs disrupts the propagation of activity to the caudal spinal cord**

(A) Profile of spontaneous fictive locomotor activity in ChR2[+] larvae recorded at rostral (VR8) and caudal (VR16) segments. The field of photoactivation was limited by narrowing the aperture of the microscope condenser.

(B) Fictive locomotor activity was silenced in the rostral VNR as well as in the caudal VNR following the activation of rostral CSF-cNs (segments 1-10). See also Figure S4.

(C) In contrast, fictive locomotor activity in the same larva was not altered following the activation of caudal CSF-cNs (segments 16-25).

(D) Effect of activation of caudal CSF-cNs versus rostral CSF-cNs. Mean bout duration calculated from VNR recordings at the segment 8 (VR 8) and segment 16 (VR 16) in $LED_{OFF}$ (black) or $LED_{ON}$ (blue) conditions.

(E) CSF-cNs-mediated silencing efficiency, calculated as the relative reduction of bout duration after CSF-cNs activation ($BD_{LED\ OFF}/BD_{LED\ ON}$ ratio), measured at segment 8 (VR 8) and 16 (VR16) during rostral or caudal activation of CSF-cNs. *, $p < 0.05$, **, $p < 0.01$ following Student's t test. Data are represented as mean ± SEM.

# Supplemental information

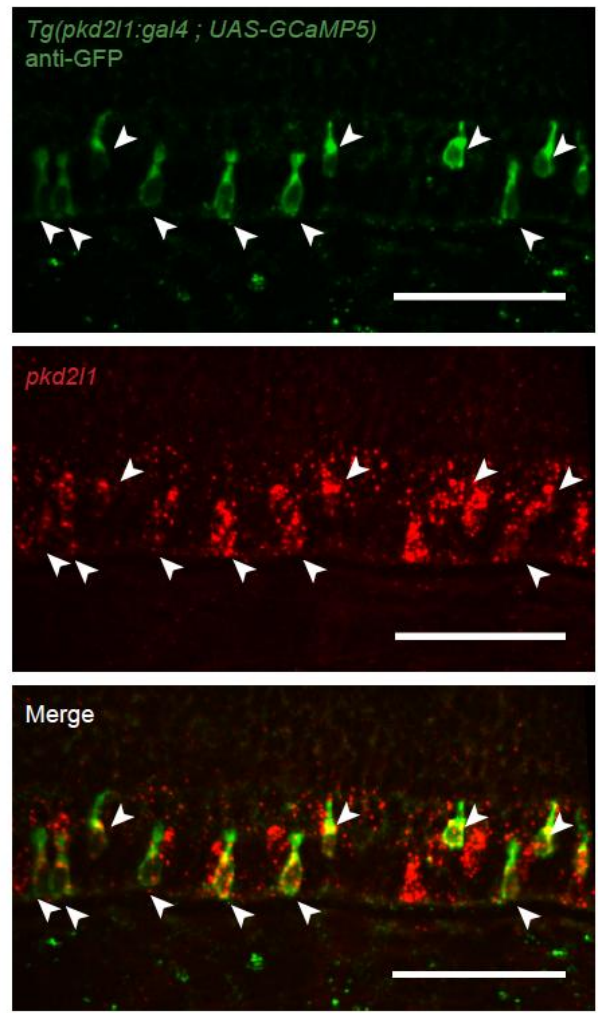

**Figure S1 related to Figure 1. The *Tg(pkd2l1:gal4)$^{icm10}$* line drives expression in *pkd2l1* positive cells in the zebrafish larva.**

Fluorescent *in situ* hybridization (FISH) for *pkd2l1* was combined with an immunohistochemistry against GFP in a *Tg(pkd2l1:gal4;UAS :GCaMP5G)* larva at 3dpf. Spinal neurons expressing GCaMP5G (green) are positive for *pkd2l1* (Pkd2l1$^+$ in red), see arrowheads, n = 5 larvae.

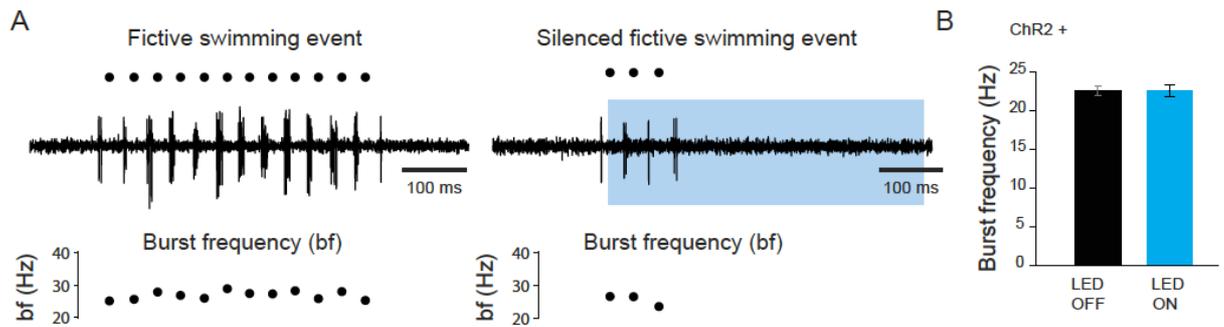

**Figure S2 related to Figure 5. The burst frequency is not affected by the activation of CSF-cNs during ongoing fictive slow locomotion.**

(A) Estimation of the burst frequency (in Hz) during spontaneous fictive slow swim bouts at baseline (LED$_{OFF}$, left) and after activation of CSF-cNs (LED$_{ON}$, right). The blue bar represents the light pulse.

(B) Quantification of the burst frequency in LED$_{OFF}$ vs LED$_{ON}$ condition in ChR2$^+$ larvae. The burst frequency is not altered by CSF-cN activation during ongoing slow swim events (n = 11 larvae).

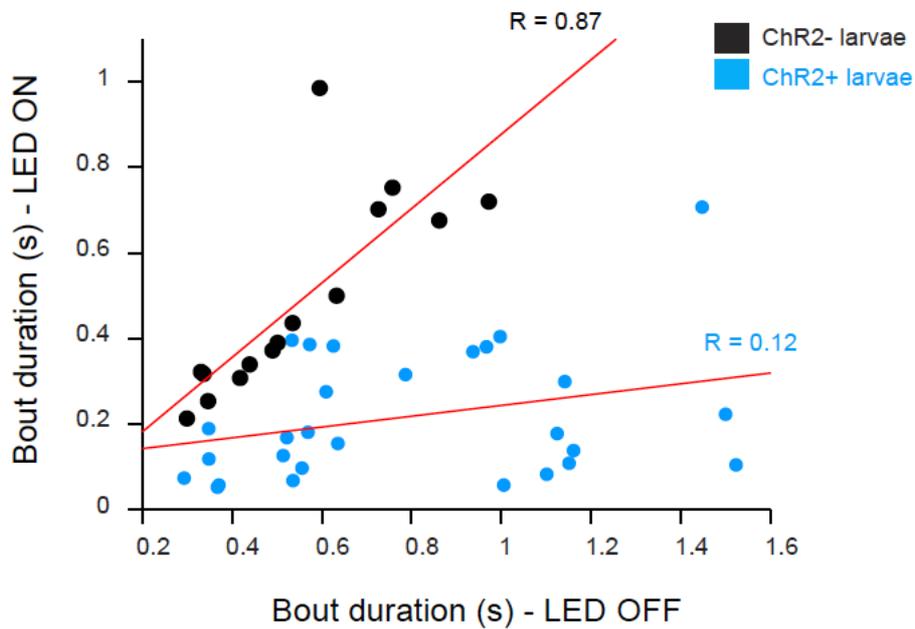

**Figure S3 related to Figure 5. Bout duration upon CSF-cN activation did not depend on the initial bout duration.**

The scatter plot shows the correlation between duration of fictive swimming events in $LED_{OFF}$ and $LED_{ON}$ conditions for $ChR2^-$ and $ChR2^+$ larvae. High correlation is observed between bout duration during $LED_{OFF}$ and $LED_{ON}$ conditions in $ChR2^-$ larvae ($p < 0.001$) but not in $ChR2^+$ larvae ($p > 0.05$). These data demonstrate that the silencing of swimming activity does not depend of the initial duration of activity in ChR2+ fish. Red lines are linear fits. R are correlation coefficients. Correlation significance was computed using Student's t test.

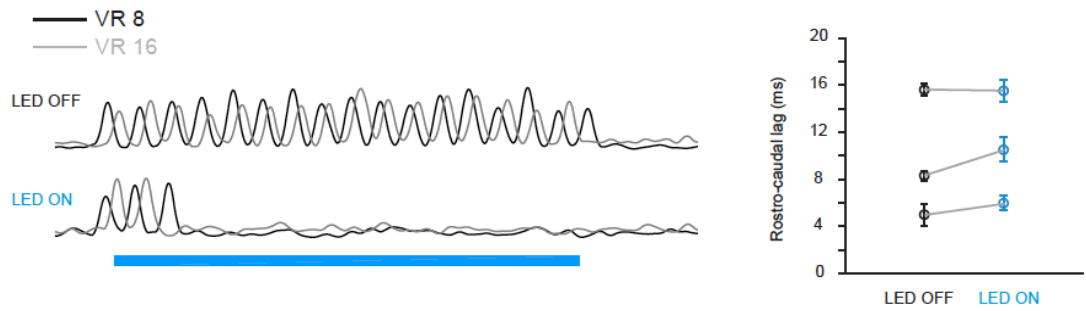

**Figure S4 related to Figure 7. The rostrocaudal lag is not affected by the activation of rostral CSF-cNs during slow locomotion.**

Enveloppe traces of VNR recordings from segment 8 and 16 shows that the caudal activity follows rostral activity by several ms illustrating that the locomotor activity propagates from rostral to caudal segments. In ChR2$^+$ larvae, the rostrocaudal lag is unaltered by the stimulation of CSF-cNs. The blue bar represents the light pulse.

**Table S1 related to Figure 1 and Figure 2. Stable transgenic lines.**

| Table S1: Transgenic lines | | | |
|---|---|---|---|
| Name | Other Name | Labelling in the spinal cord | Original publication |
| *Tg(pkd2l1:gal4)[icm10]* | - | CSF-cNs | This paper |
| *Tg(UAS:ChR2-YFP)[icm11]* | *Tg(UAS:ChR2(H134R)-eYFP)* | - | This paper |
| *Tg(gad1b:GFP)* | - | GABAergic interneurons | Satou et al., 2013 [S1] |
| *Tg(vglut2a:lox:DsRed:lox:GFP)* | - | Glutamatergic interneurons | Koyama et al., 2011 |
| *Tg(UAS:GCaMP5G)[icm08]* | *Tg(UAS:GCaMP5)* | - | This paper |

**Supplemental Experimental Procedures**

**Generation of stable transgenic lines**

To generate the *Tg(pkd2l1:gal4)$^{icm10}$* transgenic line expressing the Gal4 reporter gene in *pkd2l1* expressing cells, we amplified 3.8 kbp of genomic sequence immediately upstream of the predicted ATG site for the zebrafish *pkd2l1* gene (ENSDARG00000022503). Analyzing the genomic locus in various Teleostian species, the second intron of the predicted transcript was remarkably conserved in length. We amplified this sequence from zebrafish cDNA as it was likely to contain regulatory elements recapitulating *pkd2l1* expression. Both PCR fragments were cloned in the pCRII vector using the TOPO TA cloning kit according to manufacturer's instructions (Invitrogen). The correct DNA inserts were verified by sequencing. Both DNA fragments were then subcloned in the pT2KhspGFF plasmid [S2] using standard molecular biology techniques. The 3.8 kb promoter region was cloned upstream of the Gal4 (GFF) sequence in place of the *hsp* promoter present in the pT2KhspGFF plasmid, while the intronic sequence was placed after the Gal4 stop codon and before the SV40 poly-A site. To generate a stable transgenic line, a solution containing 20 ng/μl plasmid DNA, 50 ng/μl *tol2* transposase mRNA was injected into *Tg(UAS:Kaede)$^{s1999t}$* embryos [S3] to reveal Gal4 expression. Injected embryos were screened at 2 dpf under a dissecting fluorescent microscope. Embryos showing Gal4 expression in the spinal cord were raised and outcrossed to obtain a stable line. We generated the *Tg(UAS:ChR2-YFP)$^{icm11}$* line by injecting the original published construct UAS:ChR2(H134R)-eYFP [S4]. The *Tg(UAS:GCaMP5G)$^{icm08}$* line was generated by regular cloning in the PT2 14xUAS vector from Prof. Kawakami's lab. 50 ng/μl of this vector were injected with Tol2 RNA into eggs at the one-cell stage from the outcross of Et(e1b:Gal4-VP16)s1020t [S3] with wild type TL. Potential founders were screened later on by outcrossing to wild type animals.

**Generation of animals with mosaic labeling of CSF-cNs**

To trace individual CSF-cNs, we took advantage of multiple approaches relying on the specific *Tg(pkd2l1:Gal4)$^{icm10}$* line. For 36 cells, we injected the UAS:synaptophysin-GFP [S5] DNA construct at 60 ng/µl into single cell-stage embryos. We reconstructed 29 cells after imaging highly variegated carriers of the *Tg(pkd2l1:gal4;UAS:ChR2-mCherry)* constructs. For 22 cells, *the Tg(pkd2l1:tagRFP)* construct generated with a three-fragment Gateway recombineering reaction (Invitrogen, Carlsbad, CA, USA) was injected into wild-type embryos at 25 ng/µl. For one cell, we used the *Tg(Brn3c:Gal4:UAS:GFP)* transgenic line (BGUG) to label a subset of Gal4 expressing cells [S3, S6]. 3 dpf larvae selected for single CSF-cN expression were fixed with 4% PFA for 3 hours and immunostained using one of following antibodies: chicken anti-GFP (1:500, Abcam, Cambridge, UK), mouse anti-mCherry (1:500, Clontech, Mountain View, CA, USA) and rabbit anti-tagRFP (1:500, Evrogen, Moscow, Russia).

**Morphological analysis of isolated CSF-cNs**

Images of single CSF-cNs were acquired using an FV1000 confocal microscope (Olympus, Tokyo, Japan) equipped with a 40x water immersion objective using the 473 nm and 543 nm laser lines. For each cell, multiple z-stacks (step size 1 µm) were acquired to capture the entire arborization of the axon. Multiple stacks from a single cell were combined using the Grid/Collection stitching plugin in Fiji [S7] or the XuvTools stitching software [S8]. Cell morphology was reconstructed into a three-dimension image using the Simple Neurite Tracer (SNT) plugin in Fiji [S9]. The total axon length and branching hierarchy were obtained from SNT. Other morphological parameters were extracted using a custom-made MATLAB script. The dorsoventral (D-V) soma positions were measured from the center of the cell body and normalized to the limits of the spinal cord. Counting of presynaptic boutons was performed in a semi-automated manner using a custom-made MATLAB script. To distinguish boutons from vesicles, we imaged live segments of CSF-cNs axons (n = 23, Movie S1). We noticed that vesicles were recognizable as they were small and rapidly migrating along the axon while boutons were bright and immotile (Movie S1).

**Fluorescent *in situ* hybridization (FISH) and immunohistochemistry (IHC)**

The procedure used here relied on published protocols [S10]. Briefly, whole-mount ISH were performed on 3 dpf larvae fixed in 4% PFA in PBS overnight at 4°C. To reveal *pkd2l1* expression, probes [S10] were detected with anti-DIG antibodies (Roche Diagnostics, France). *pkd2l1* FISH was performed before IHC against GFP: embryos were washed and immunostained with the chicken anti-GFP antibody antibody overnight at 4°C, and then incubated with the corresponding Alexa conjugated secondary antibodies IgG (1:500, Life Technologies) combined with DAPI (2.5 µg/ml, Life Technologies).